\documentclass[twocolumn,showpacs,preprintnumbers,amsmath,amssymb,nofootinbib]{revtex4}

\input psfig.sty
\usepackage{epsfig}

\usepackage{graphicx}
\usepackage{dcolumn}
\usepackage{bm}

\newcommand{\be}{\begin{equation}}
\newcommand{\ee}{\end{equation}}
\begin{document}

\title{Cosmological consequences of a possible $\Lambda$-dark matter interaction}

\author{F. E. M. Costa}\email{ernandes@on.br}

\author{J. S. Alcaniz}\email{alcaniz@on.br}

\affiliation{Observat\'orio Nacional, 20921-400, Rio de Janeiro -- RJ, Brasil}

\date{\today}

\begin{abstract}
We propose a general class of interacting models in which the interaction between the CDM component and $\Lambda$ is parameterized by an arbitrary function of the cosmic scale factor  $\epsilon(a)$. Differently from other dynamical $\Lambda$ scenarios in which the final stage of cosmic expansion is a de Sitter phase, we find solutions of transient acceleration, in which the $\Lambda$-dark matter interaction will drive the Universe to a new dark matter-dominated era in the future. We investigate some cosmological consequences of this model and discuss some constraints on its parameters from current SNe Ia, BAO, CMB and $H_0$ data.

\end{abstract}


\maketitle

\section{Introduction}

According to current observational results, a mysterious field, named dark energy, which accounts for $\sim 70\%$ of the cosmic composition, is governing the late time dynamics of the Universe. However, although fundamental to our understanding of the Universe, several important questions about the nature of this dark energy component and its role in the cosmic dynamics remain unanswered (see, e.g., \cite{review} for some recent reviews).

Among many possible candidates, perhaps the simplest explanation for current observations is that the unclumped form of energy density corresponds to a positive cosmological constant $\Lambda$, whose presence modifies the Einstein field equations to  
\begin{equation} \label{efe}
G^{\mu \nu} = \chi T^{\mu \nu} + {\Lambda g^{\mu \nu}}\;,
\end{equation}
where $G^{\mu \nu}$ is the Einstein tensor, $T^{\mu \nu}$ is the energy-momentum tensor of matter fields and CDM particles, and $\chi = 8\pi G$ is the Einstein's constant (throughout this paper we work in units where the speed of light $c = 1$).

From the observational point of view, it is well known that flat models with a very small cosmological term ($\rho_{\Lambda} \lesssim 10^{-47}$ ${\rm{GeV}}^4$) are in good agreement with almost all sets of cosmological observations, which makes them an excellent description of the observed Universe. From the theoretical viewpoint, however, at least two problems still remain. First, and possibly the most serious one is the unsettled situation in the particle physics/cosmology interface (the so-called cosmological constant problem (CCP)~\cite{weinberg}), in which the cosmological upper bound differs from theoretical expectations ($\rho_{\Lambda} \sim 10^{71}$ ${\rm{GeV}}^4$) by more than 100 orders of magnitude. The second is that, although a very small (but non-zero) value for $\Lambda$ could conceivably be explained by some unknown physical symmetry being broken by a small amount, one should be able to explain not only why it is so small but also why it is exactly the right value that is just beginning to dominate the energy density of the Universe now. Since both components (dark matter and dark energy) are usually assumed to be independent and, therefore, scale in different ways, this would require an unbelievable coincidence, the so-called coincidence problem (CP).

A phenomenological attempt at alleviating the CP problem is allowing the dark matter and dark energy to interact (in the case of a $\Lambda$-dark matter interaction, $\Lambda$ is necessarily a time-dependent quantity, which is the basic idea behind the decaying $\Lambda$ models whose aim is to solve or alleviate the CCP)\footnote{Strictly speaking, in the context of classical general relativity any additional $\Lambda$-type term that varies in space or time should be thought of as a new time-varying field and not as a cosmological constant. Here, however, we adopt the usual nomenclature of time-varying or dynamical $\Lambda$ models.}. Cosmological scenarios with a dynamical $\Lambda$ term were independently proposed about two decades ago in Ref.~\cite{list} (see also \cite{list2,friedman,shapiro,wm,alclim05,ernandes,saulo}) whereas models of couped quintessence (in which the dark energy is represented by a scalar field $\phi$ or by a smooth component parameterized by an equation of state $p_{\rm{DE}} = w \rho_{\rm{DE}}$ with $w < 0$) have been investigated more recently~\cite{cq,jesus,ernandes1}. In both cases, however, the absence of a natural guidance from fundamental physics on a possible interacting or coupling term between the two dark components leads most of the current investigations discussed in the literature to a  phenomenological level. 

In this regard, a still phenomenological but very interesting step toward a more realistic interacting or coupling law was recently discussed in Ref.~\cite{wm}, in which the time dependence of the $\Lambda$ term is deduced from its effect on the CDM evolution. Such a coupling is similar to the one obtained in Ref.~\cite{shapiro} from arguments based on renormalization group and seems to be very general, having many of the previous attempts as a particular case. Thermodynamical considerations for this class of dynamical $\Lambda$ models~\cite{alclim05} showed that the interacting parameter $\epsilon$ must be positive, which means that the energy transfer between $\Lambda$ and the dark matter field is such that the latter always gains energy from the former, and not the other way around. In Refs.~\cite{ernandes}, a scalar field description for this class of dynamical $\Lambda$ models was investigated and found to be well represented by a coupled double exponential potential of the type $V(\Phi) \propto \exp{(\lambda \phi)} + \exp{(-\lambda \phi)}$. In Ref.~\cite{jesus}, a coupled quintessence model based on the above arguments (for which the equation-of-state parameter $w \neq -1$) as well as some observational constraints on the interacting term were discussed.

An important aspect worth emphasizing is that in the above analyses the interacting parameter $\epsilon$ has been considered constant over the cosmic evolution whereas in a more realistic case it must be a time-dependent quantity. Our goal in this paper is therefore to go a step further in the above description and extend the arguments of Refs.~\cite{wm,alclim05,ernandes} to a physically more realistic case in which $\epsilon$ is a function of time. We restrict the present analysis to coupled quintessence models in which $w_\phi = -1$, which is mathematically equivalent to dynamical $\Lambda$ scenarios. In terms of the interacting function $\epsilon(a)$, we discuss the dynamical behavior of this class of models and find viable cosmological solutions for a subset of values of $\epsilon(a)$. The possibility of transient accelerating solutions in which the Universe will experience a future dark matter-dominated phase is also explored. We also carry out a joint statistical analysis with recent observations of SNe Ia, BAO, CMB and $H_0$ to check the observational viability of this general class of interacting $\Lambda$-dark matter models.

\section{The model: Basic equations}

From Eq.~(\ref{efe}) the Bianchi identities imply that the coupling between a dynamical $\Lambda$ term and CDM particles must be of the type\footnote{In our analysis, we will consider only interaction between $\Lambda$ and CDM particles. For a discussion about bounds on the interaction with conventional matter from local gravity experiments, see~\cite{pdg}. For constraints from primordial nucleosynthesis, see also~\cite{friedman}.} 
\begin{equation} \label{lt}
u_\mu {\cal{T}}^{\mu \nu};_{\nu} = -u_\mu\left(\frac{\Lambda g^{\mu \nu}}{\chi}\right);_{\nu}\;,
\end{equation}
or, equivalently,
\begin{equation} \label{ec}
\dot{\rho}_{dm} + 3\frac{\dot{a}}{a}\rho_{dm} = - \dot{\rho}_\Lambda\;, 
\end{equation}
where $\rho_{dm}$ and $\rho_\Lambda$ are the energy densities of CDM and $\Lambda$, respectively, and ${\cal{T}}^{\mu \nu} = \rho_{dm} u^{\mu}u^{\nu}$ stands for the energy-momentum tensor of the CDM field. As usual, the dot sign denotes derivative with respect to the time.

The $\Lambda$-CDM interaction implies that the energy density of this latter component must dilute at a different rate compared to its standard evolution, $\rho_{dm} \propto a^{-3}$, where $a$ is the cosmological scale factor. Thus, the deviation from the standard dilution may be characterized by the function $\epsilon(a)$, such that
\begin{equation} \label{dm}
\rho_{dm} = \rho_{dm, 0}a^{-3 + \epsilon(a)}\;,
\end{equation}
where we have set the present-day value of the cosmological scale factor $a_0 = 1$.  Since the other matter fields (radiation and baryons) are separately conserved, Eqs. (\ref{ec}) and (\ref{dm}) provide
\begin{equation}\label{decayv}
\rho_{\Lambda} =  \rho_{dm,0} \int_{a}^{1}{\epsilon(\tilde{a}) + \tilde{a}\epsilon' \ln(\tilde{a}) \over \tilde{a}^{4 - \epsilon(a)}} d\tilde{a} + {\rm{X}}\;,
\end{equation}
where a prime denotes derivative with respect to the scale factor and ${\rm{X}}$ is an integration constant.

\begin{figure*}
\centerline{\psfig{figure=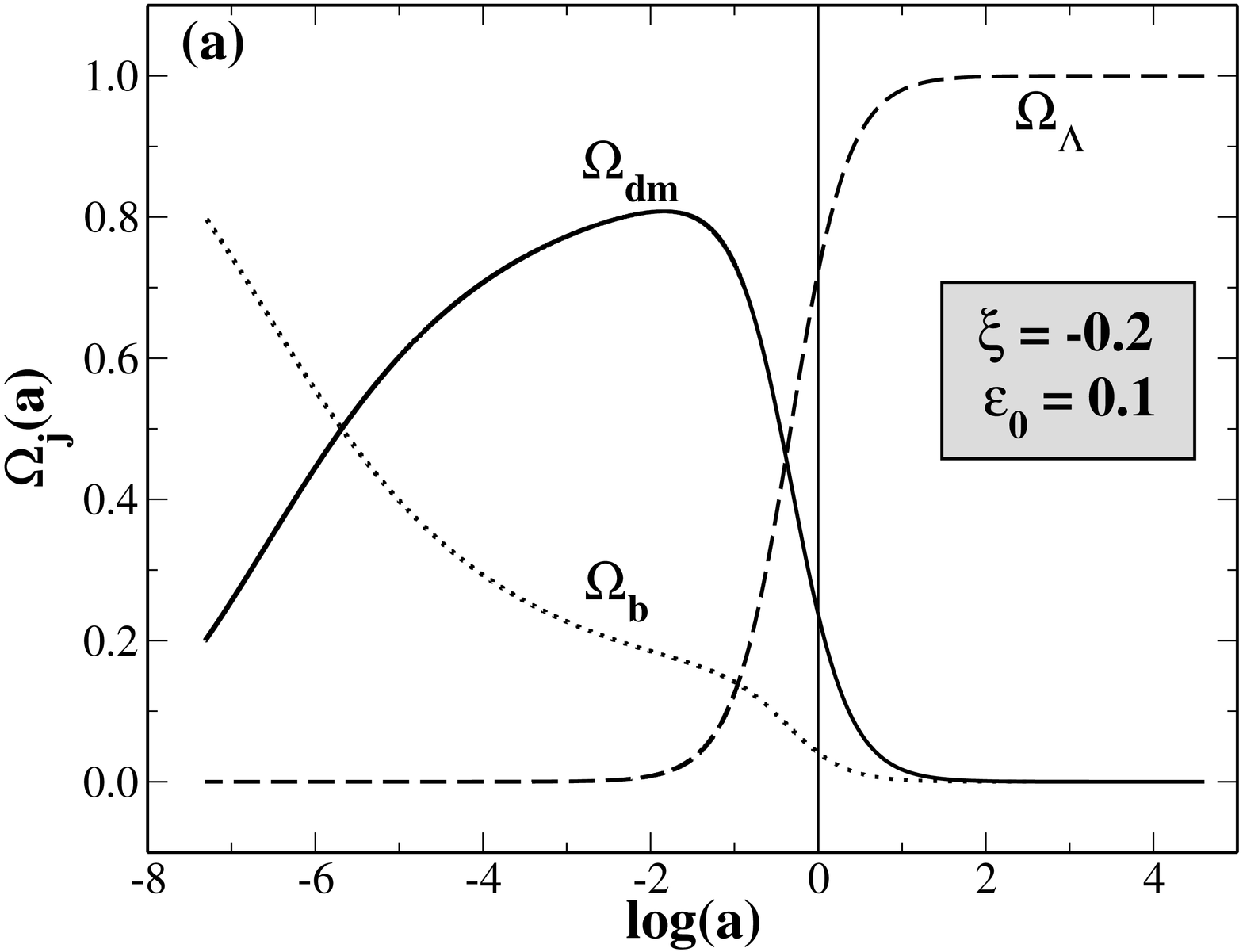,width=2.2truein,height=2.4truein,angle=-90} 
\psfig{figure=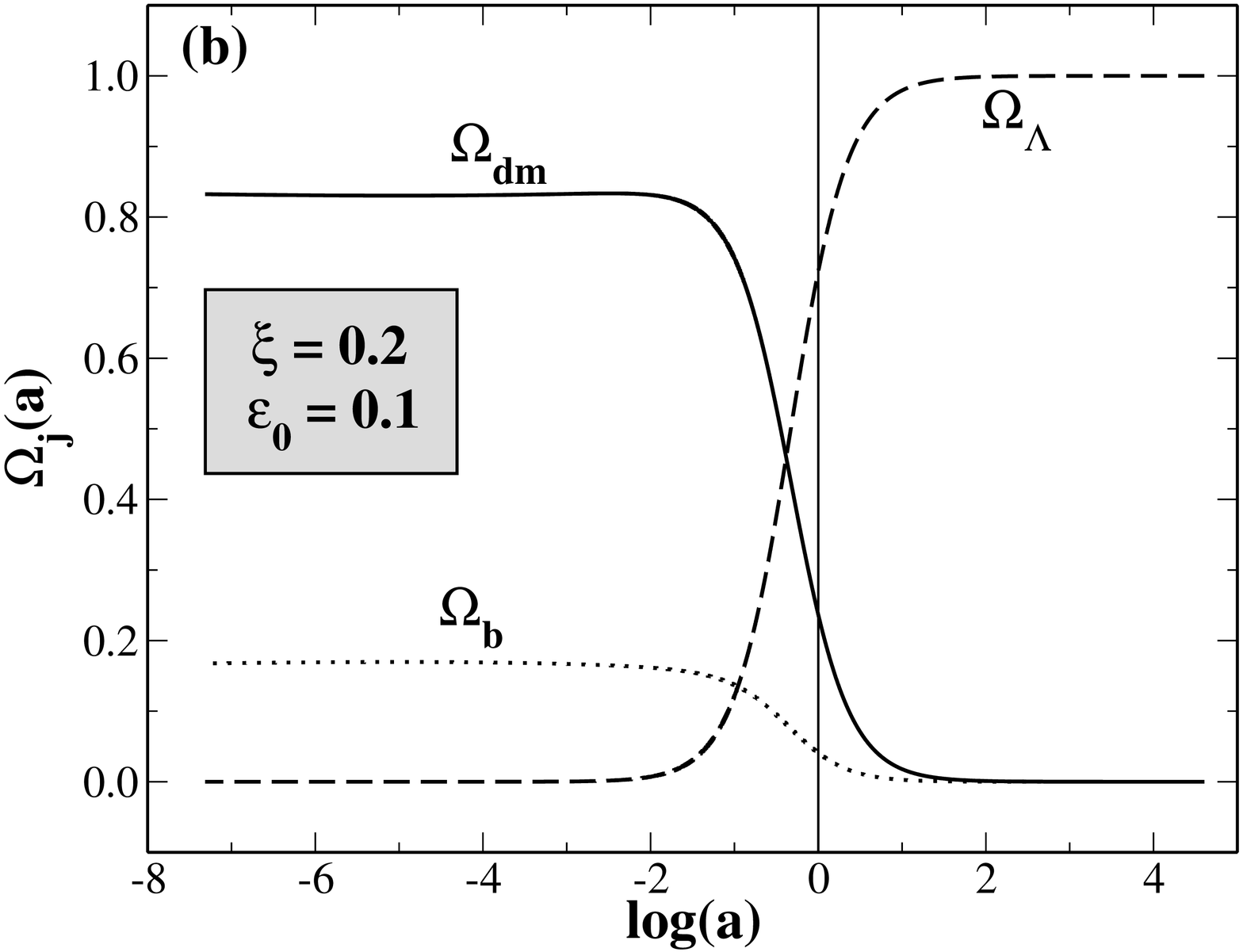,width=2.2truein,height=2.4truein,angle=-90}
\psfig{figure=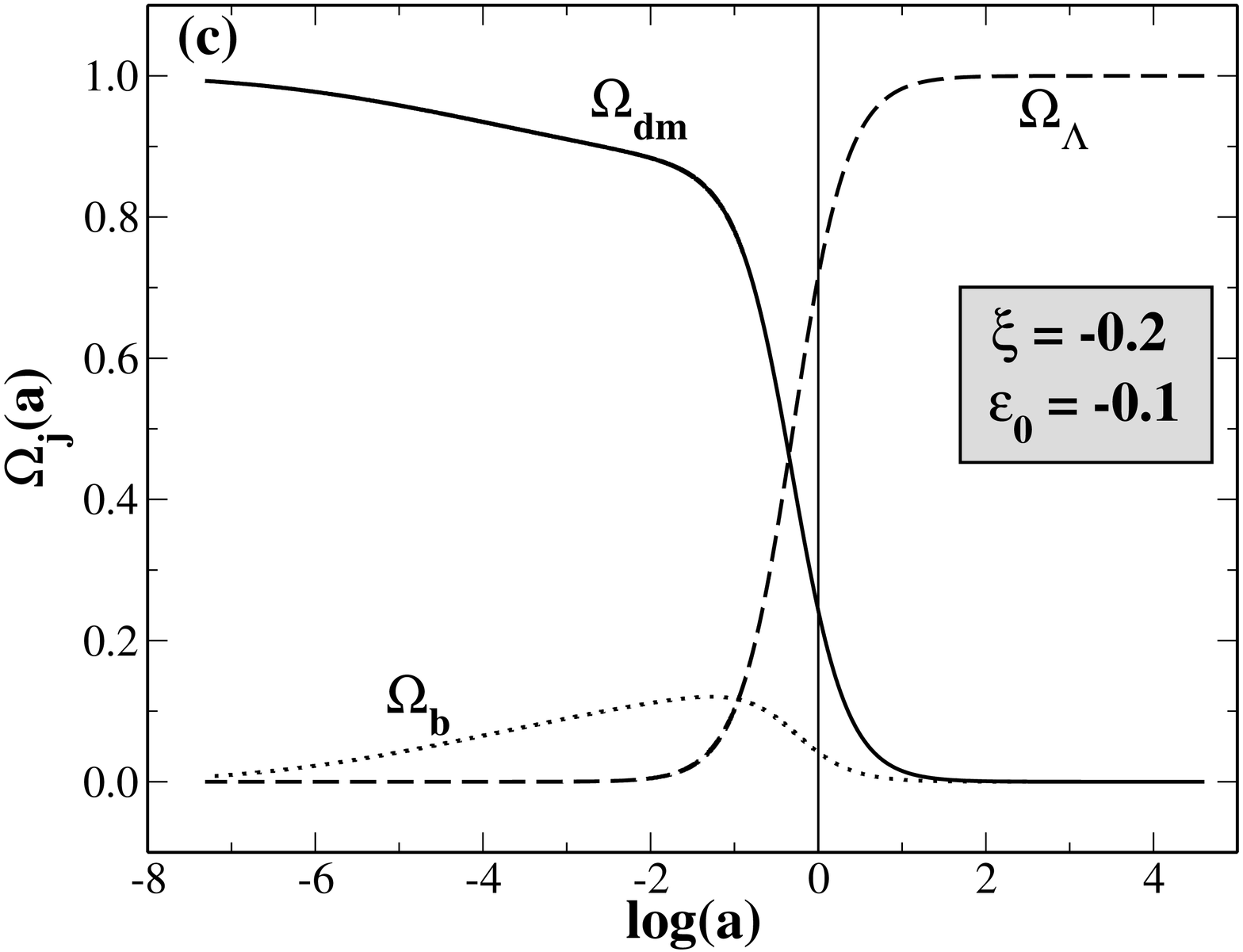,width=2.2truein,height=2.4truein,angle=-90}} 
\caption{Evolution of the density parameters $\Omega_j$ ($j = b, dm, {{\rm{X}}}$) as a function of $\log(a)$ for some selected combinations of $\epsilon_0 = \pm 0.1$ and $\xi = \pm 0.2$ and ${\rm{A}} \simeq 17.27$, ${\rm{B}} = 0.17$ and  ${\rm{D}} \simeq 3$.}
\label{fig:qzw}
\end{figure*}

\begin{figure*}
\centerline{\psfig{figure=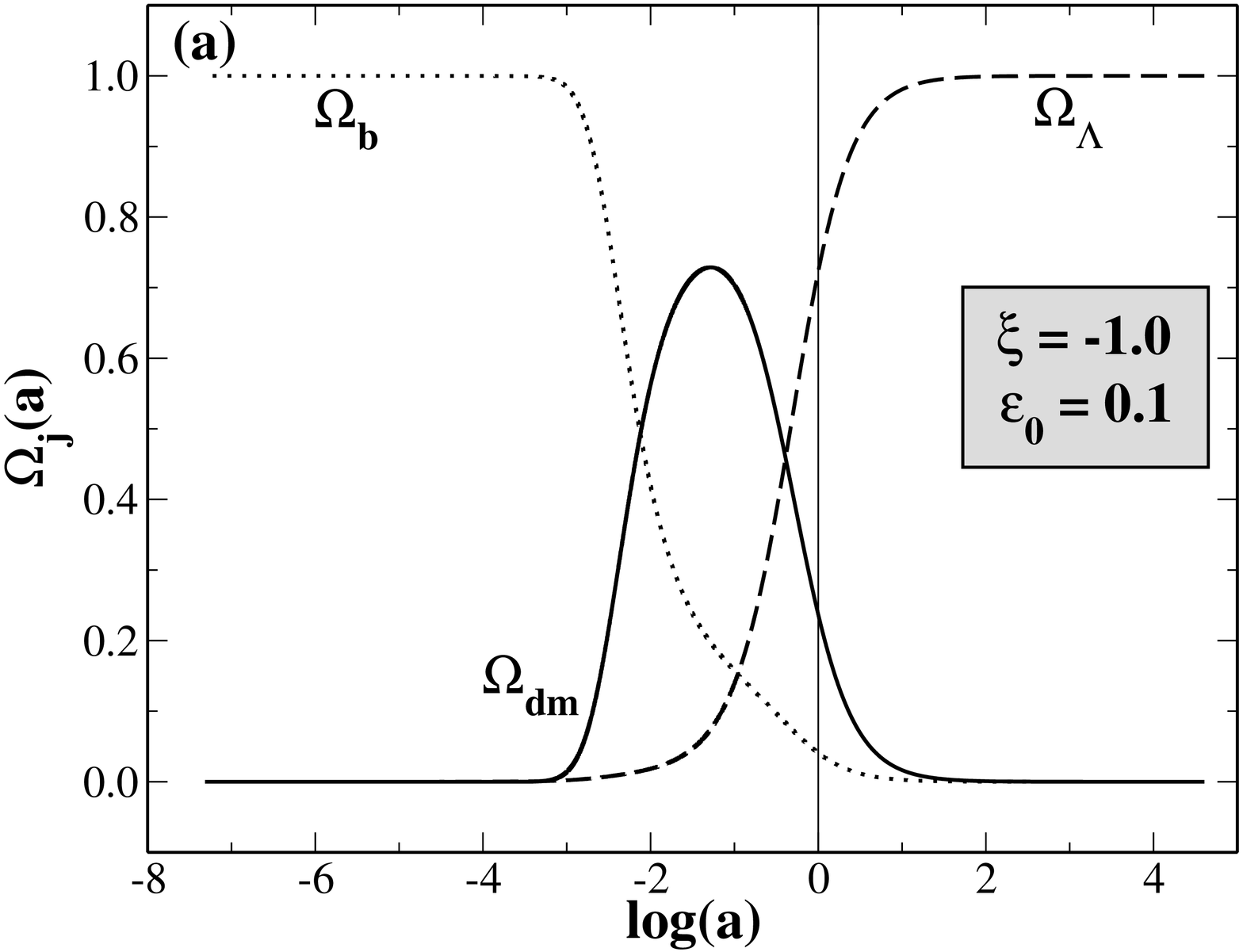,width=2.2truein,height=2.4truein,angle=-90} 
\psfig{figure=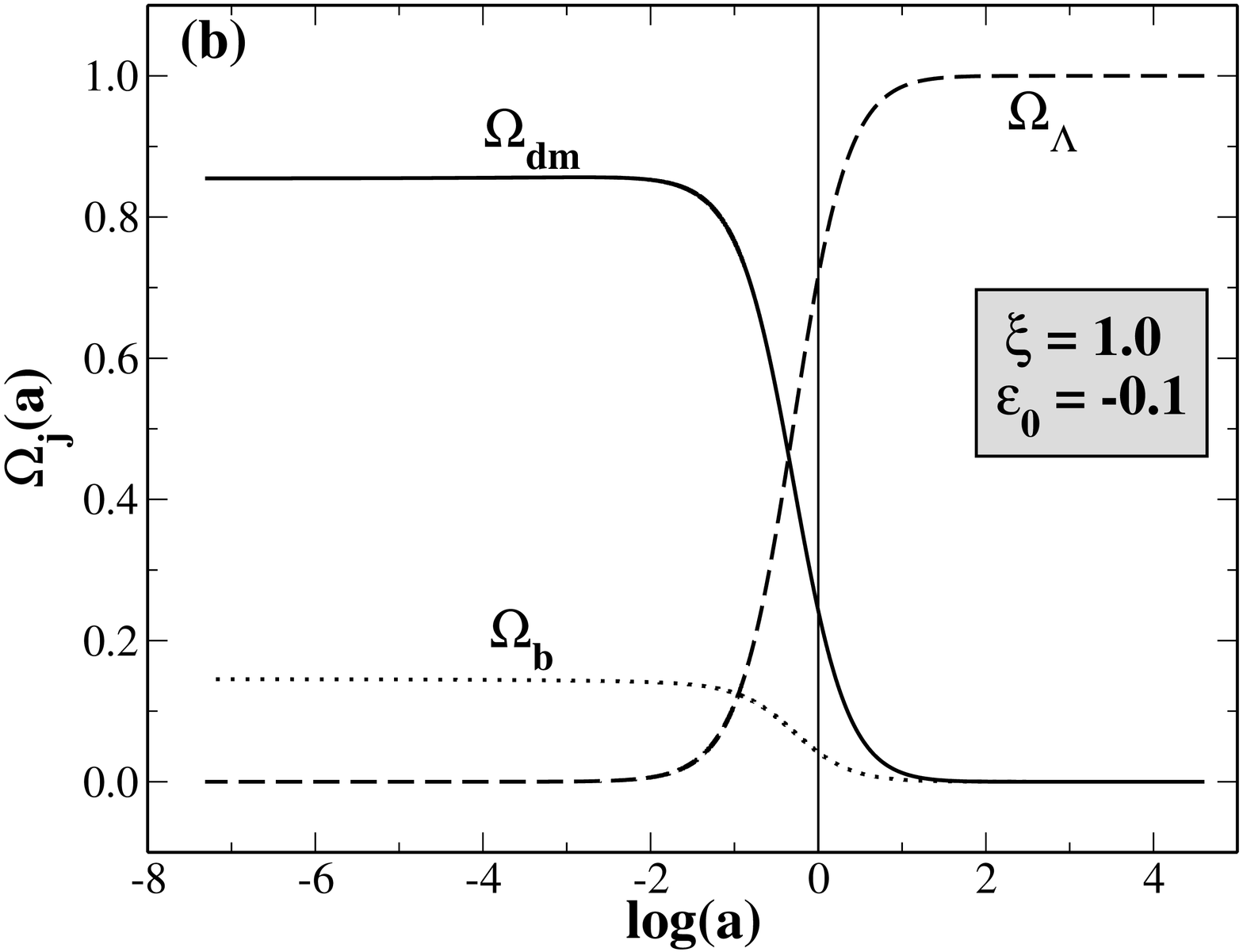,width=2.2truein,height=2.4truein,angle=-90}
\psfig{figure=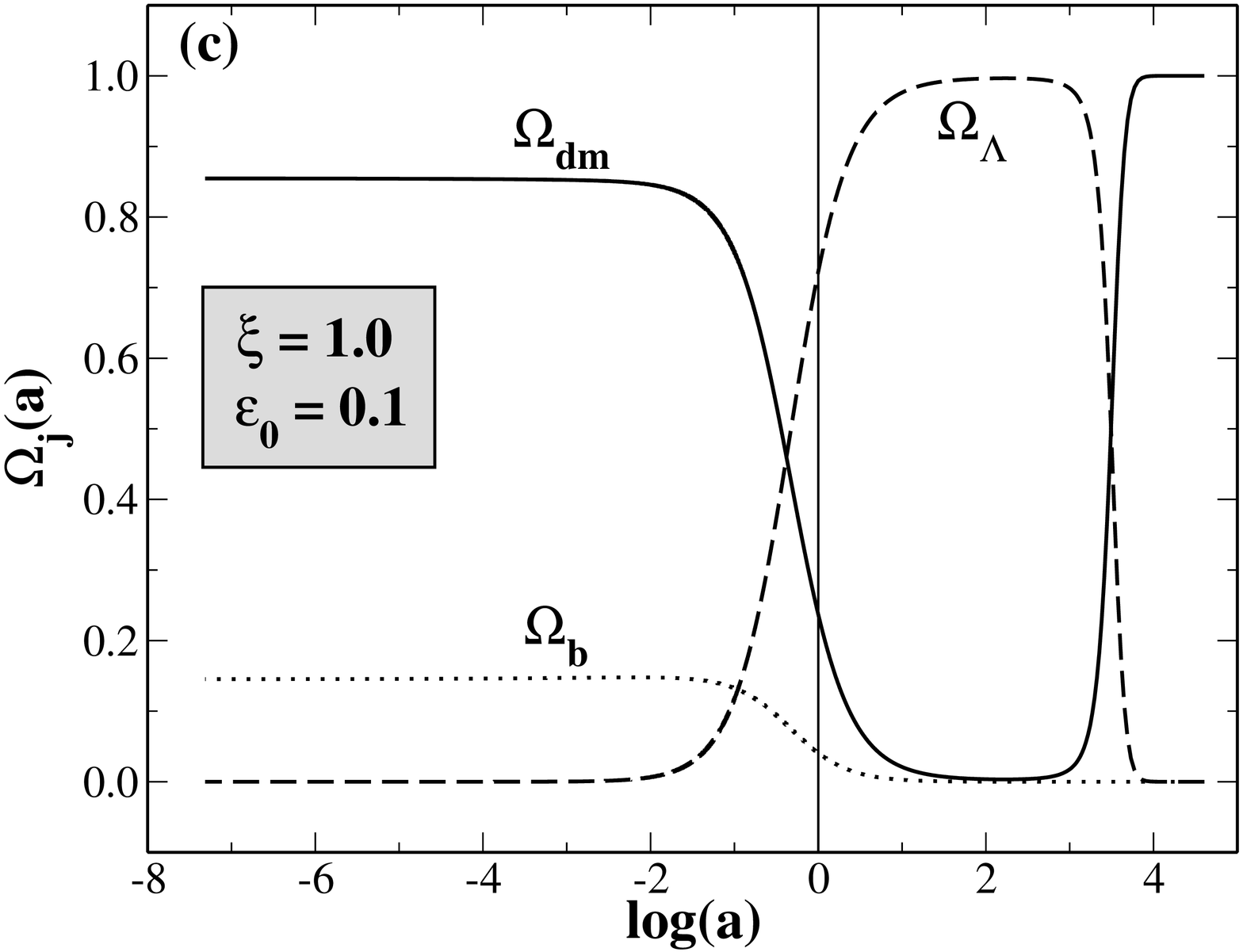,width=2.2truein,height=2.4truein,angle=-90}} 
\caption{The same as in Fig. 1 for combinations of $\epsilon_0 = \pm 0.1$ and $\xi = \pm 1.0$. Note that for $\epsilon_0 > 0$ and large positive values of $\xi$ (Panel 1c), the $\Lambda$-dark matter interaction will drive the Universe to an eternal deceleration instead of the usual de Sitter phase.}
\label{fig:qzw}
\end{figure*}

Neglecting the radiation contribution, the Friedmann equation for this dynamical $\Lambda$ cosmology can be rewritten as
\begin{equation}
\label{friedmann} {{H}}= H_0\left[\Omega_{b,0}{a}^{-3} + \Omega_{dm,0}\varphi(a) + {\Omega}_{{\rm{X,0}}}\right]^{1/2},
\end{equation}
where $H(z)$, $\Omega_{b,0}$ and $\Omega_{dm,0}$ are, respectively, the Hubble parameter, and the baryons and CDM present-day density parameters. The function $\varphi(a)$ is written as 
\begin{equation} \label{f(a)}
\varphi(a) = a^{-3 +\epsilon(a)} + \int_{a}^{1}{\epsilon(\tilde{a}) + \tilde{a}\epsilon' \ln(\tilde{a}) \over \tilde{a}^{4 - \epsilon(a)}} d\tilde{a}\;,
\end{equation}
and ${\Omega}_{{\rm{X,0}}}$ stands for the present-day relative contribution of the constant ${\rm{X}}$ to the expansion rate.

In order to proceed further, we must assume an appropriated relation for $\epsilon(a)$. Certainly, among many possible functional forms, a very simple choice is 
\begin{eqnarray}
\label{Parametrization_a}
\epsilon(a)& = & \epsilon_0a^\xi\quad  \nonumber \\
& = &\epsilon_0(1+z)^{-\xi} \;,
\end{eqnarray}
where $\epsilon_0$ and $\xi$ may, in principle, take negative and positive values. With the above expression, Eq. (\ref{decayv}) can be rewritten as
\begin{equation}\label{decayv2}
\rho_{\Lambda} =  \rho_{m0}\epsilon_{0} \int_{a}^{1}{[1 + \ln(\tilde{a}^{\xi})] \over \tilde{a}^{4 - \xi - \epsilon_{0}\tilde{a}^{\xi}}} d\tilde{a} + {\rm{X}}\;.
\end{equation}
Note that, in the absence of a coupling with the CDM component, i.e., $\epsilon_0 = 0$, we may identify ${\rm{X}} \equiv {\rho}_{\Lambda 0}$ (the current value of the vacuum contribution) and the standard $\Lambda$CDM scenario is fully recovered. Note also that, for $\xi = 0$ and $\epsilon_0 \neq 0$, the above expressions reduce to the dynamical $\Lambda$ scenario recently discussed in Refs.~\cite{wm,alclim05,ernandes}, whose vacuum energy density is given by
\begin{equation}
\rho_{\Lambda} =  \frac{\epsilon_0 \rho_{dm,0}}{3 - \epsilon_0}a^{-3 + \epsilon_0} + {\bar{\rm{X}}}\;.
\end{equation}

\section{Cosmic evolution}

The time evolution of the density parameters $\Omega_b(a)$, $\Omega_{dm}(a)$ and $\Omega_{\Lambda}(a)$ (the relative contribution of $\rho_\Lambda$ [Eq. \ref{decayv2}] to the expansion rate) can be derived by combining Eqs. (\ref{dm})-(\ref{friedmann}). They read
\begin{subequations}
\begin{equation} \label{8a}
\Omega_{b}(a) = \frac{a^{-3}}{{\rm{A}} + a^{-3} + {\rm{B^{-1}}}\varphi(a)}\;,
\end{equation}
\begin{equation} \label{8b}
\Omega_{dm}(a) = \frac{a^{-3 + \epsilon(a)}}{{\rm{D}} + {\rm{B}}a^{-3} + \varphi(a)}\;,
\end{equation}
\begin{equation} \label{8c}
\Omega_{{\rm{\Lambda}}}(a) = \frac{{\rm{D}} + \varphi(a) - a^{-3 + \epsilon(a)}}{{\rm{D}} + {\rm{B}}a^{-3}  + \varphi(a)}\;,
\end{equation}
\end{subequations}
where  ${\rm{A}} = {\Omega_{{\rm{X}},0}}/{\Omega_{b,0}}$, ${\rm{B}} = {\Omega_{b,0}}/{\Omega_{dm,0}}$ and  ${\rm{D}} = {\Omega_{{\rm{X}},0}}/{\Omega_{dm,0}}$. 

Figures 1 and 2 show the evolution of the density parameters $\Omega_{{j}}$ ($j = b, dm, \Lambda$) with the logarithm of the scale factor $\log(a)$ [Eqs. (\ref{8a})-(\ref{8c})] for values of ${\rm{A}} \simeq 17.27$, ${\rm{B}} = 0.17$ and  ${\rm{D}} \simeq 3$ (corresponding to $\Omega_{b,0} = 0.0416$ and $\Omega_{dm,0} = 0.24$) 
and $\epsilon_0 = \pm 0.1$. Two symmetric values of $\xi$, i.e., $\xi = \pm 0.2$ (Fig. 1) and $\xi = \pm 1.0$ (Fig. 2) are considered. Although currently accelerated (and, therefore, possibly in agreement with SNe Ia data), models with $\epsilon_0 > 0$ and negative values of $\xi$  (Figs. 1a and 2a) fail to reproduce the past dark matter-dominated epoch, whose existence is fundamental for the structure formation process to take place. In both cases, the dark energy and dark matter densities vanish at high-$z$ and the Universe is fully dominated by the baryons. Note that the same is not true when both $\xi$ and $\epsilon_0$ take negative values (Fig. 1c). In this case, the negative signs compensate each other so that the dark matter dominates the past evolution of the Universe while the baryonic and dark energy densities vanish.

Regardless of the sign of $\epsilon_0$, well-behaved scenarios are obtained when $\xi$ takes positive values (Figs. 1b and 2b). Note that in these cases  a mix of baryons ($\lesssim 20\%$) and dark matter ($\gtrsim 80\%$) dominates the past evolution of the Universe whereas the dark energy is always the dominant component from a value of $a_* \lesssim 1$ on. 
A very interesting and completely different future cosmic evolution is obtained when $\epsilon_0 > 0$ and the parameter $\xi$ takes large positive values ($\gtrsim 0.8$). This is shown in Fig. 2c for $\xi = 1.0$ and $\epsilon_0 = 0.1$. Note that, besides having a well-behaved past evolution and being currently accelerating, the cosmic acceleration will eventually stop at some value of $a>>1$ (when the dark energy becomes sub-dominant) and the Universe will experience a new matter-dominated era in the future, when $a \rightarrow \infty$. This kind of dynamic behavior is not  found in most of the dynamical $\Lambda$ or coupled quintessence models discussed in the literature, being essentially a feature of the so-called thawing~\cite{thaw} and hybrid~\cite{cqg} potentials, which in turn seems to be in good agreement with some requirements of String or M theories, as discussed in Ref.~\cite{fischler} (see also \cite{ed})~\footnote{The argument presented in Ref.~\cite{fischler} is that an eternally accelerating universe, a rather generic feature of many quintessence scenarios (including the standard $\Lambda$CDM model), seems not to be in agreement with String/M-theory predictions, since it is endowed with a cosmological event horizon which prevents the construction of a conventional S-matrix describing particle interactions.}.

\begin{figure}[t]
\centerline{\psfig{figure=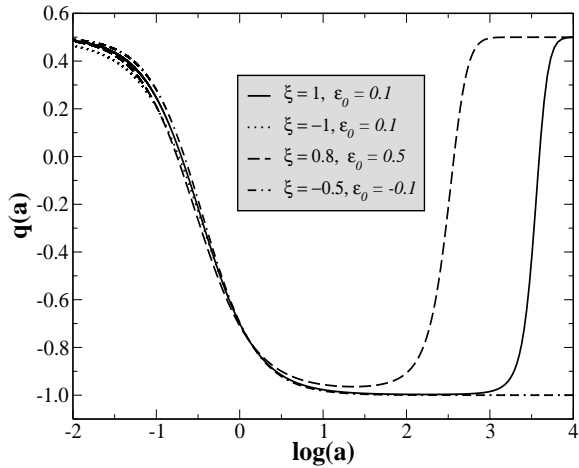,width=2.5truein,height=3.1truein,angle=-90}} 
\caption{Deceleration parameter as a function of $\log(a)$ for some selected values of $\epsilon_0$ and $\xi$. In agreement with the results shown in Figs. 1 and 2, note that for large positive values of $\xi$ the Universe will experience a new matter-dominated era in the future, when $a \rightarrow \infty$.}
\label{fig:qzw}
\end{figure}

To better visualize this transient acceleration phenomenon, we derive the deceleration parameter $q=-a\ddot{a}/\dot{a}^2$, given by 

\begin{equation}
q(a) = \frac{3}{2}\frac{\Omega_{b,0}a^{-3} + \Omega_{dm,0}a^{\epsilon(a) - 3}}{\Omega_{b,0}{a}^{-3} + \Omega_{dm,0}\varphi(a) + {\Omega}_{\rm{X},0}} -1,
\end{equation}
and shown in Fig. 3 as a function of $\log(a)$ for some selected values of $\xi$ and $\epsilon_0$. Note that for large positive values of $\xi$ the Universe was matter-dominated in the past [$q(a) \rightarrow 1/2$ for $a << 1$], switched to a long period of cosmic acceleration at $a_{acc} < 1$ but will eventually decelerate again at some $a_{dec} > 1$ (see also \cite{prl2006} for a discussion on quintessence and brane-world models of transient acceleration). 

\begin{figure}[t]
\centerline{\psfig{figure=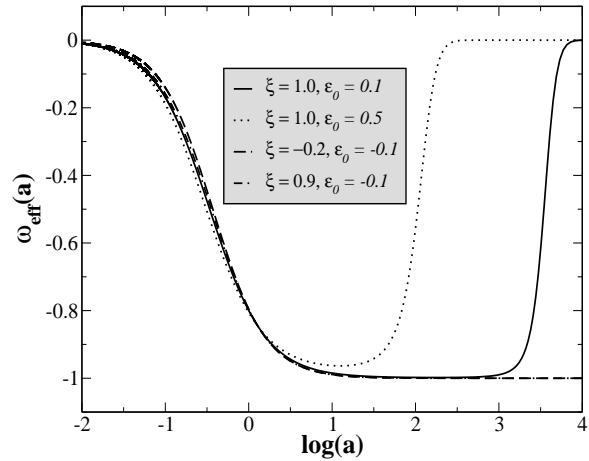,width=2.5truein,height=3.1truein,angle=-90}} 
\caption{The effective EoS $w_{eff}(a) \equiv p_{T}(a)/\rho_{T}(a)$ as a function of $\log(a)$ for some selected values of $\epsilon_0$ and $\xi$. Depending on the combination of values for $\epsilon_0$ and $\xi$ $w_{eff}$ may bahave as freezing or hybrid.}
\label{fig:qzw}
\end{figure}

For the sake of completeness, we also show in Fig. 4 the effective equation-of-state (EoS) parameter [$w_{eff}(a) \equiv p_{T}(a)/\rho_{T}(a)$] 
\begin{equation}
w_{eff}(a)  = -1 + \frac{\Omega_{b,0}a^{-3} + \Omega_{dm,0}a^{\epsilon(a) - 3}}{\Omega_{b,0}{a}^{-3} + \Omega_{dm,0}\varphi(a) + {\Omega}_{\rm{X},0}},
\end{equation}
as a function of $\log(a)$ for some combinations of $\xi$ and $\epsilon_0$. Note that, while the value of $\xi$ determines the general behavior of $w(a)$, the interacting parameter $\epsilon_0$ is directly related to the duration of the accelerating phase. Note also that, although presenting many different behaviors, clearly a very interesting one is provided by values of $\xi \simeq 1$, in which $w(a)$ behaves initially as freezing over all the past cosmic evolution, is approaching $-1$ today (in agreement with current observations), will become thawing in the near future and will behave as such  over the entire future evolution of the Universe. This freezing/thawing or hybrid behavior, originally discussed in Ref.~\cite{cqg}, is particularly interesting because, in principle, it could reconcile the slight preference of the SNe Ia and large scale structure data for freezing EoS pointed out in Refs.~\cite{krauss,trotta,huterer} (which in turn leads to an eternally accelerating Universe) with the String/M-theory requirements discussed in Ref.~\cite{fischler}.

\begin{figure*}
\centerline{\psfig{figure=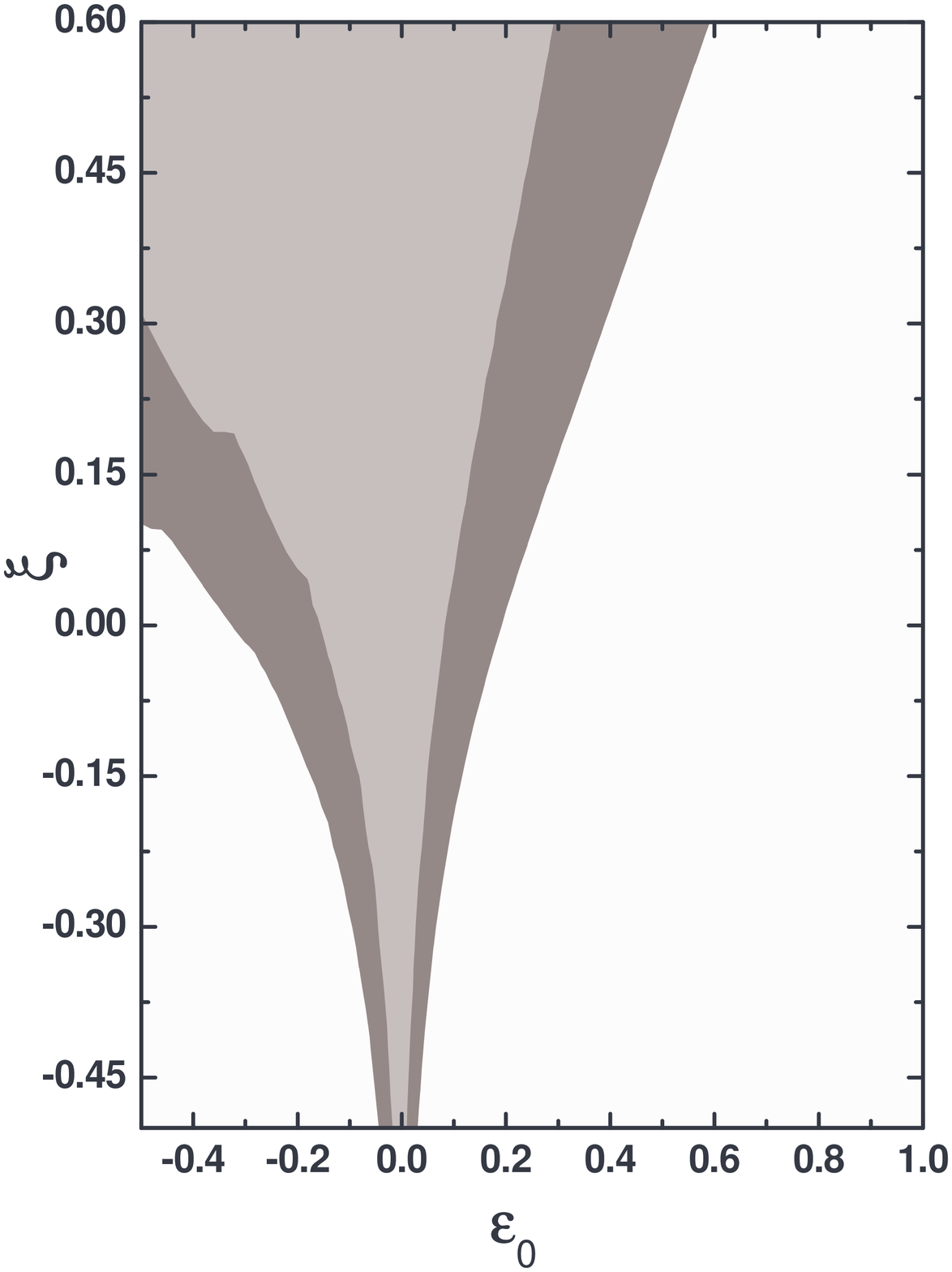,width=2.4truein,height=2.4truein,angle=0} 
\psfig{figure=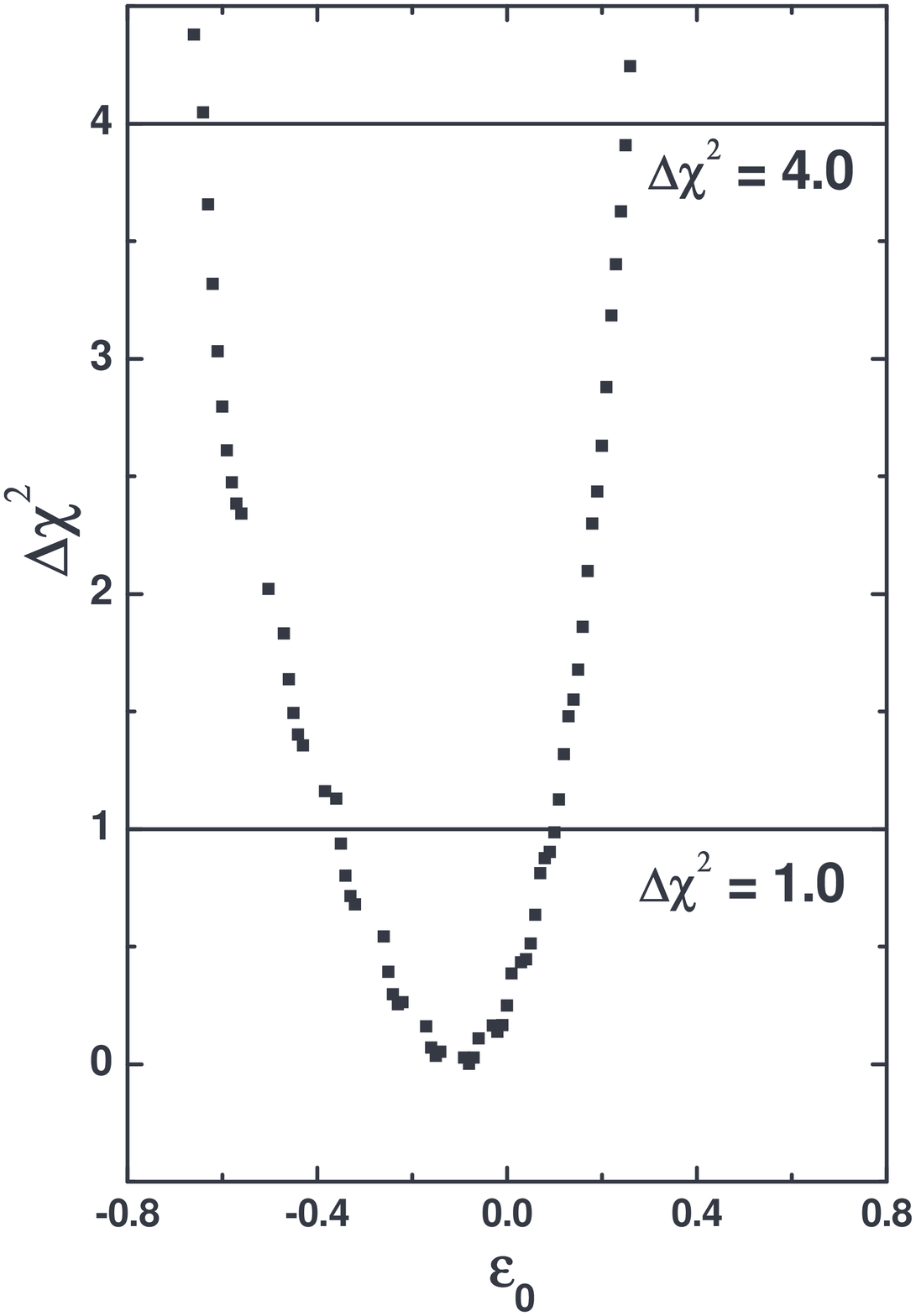,width=2.4truein,height=2.4truein,angle=0}
\psfig{figure=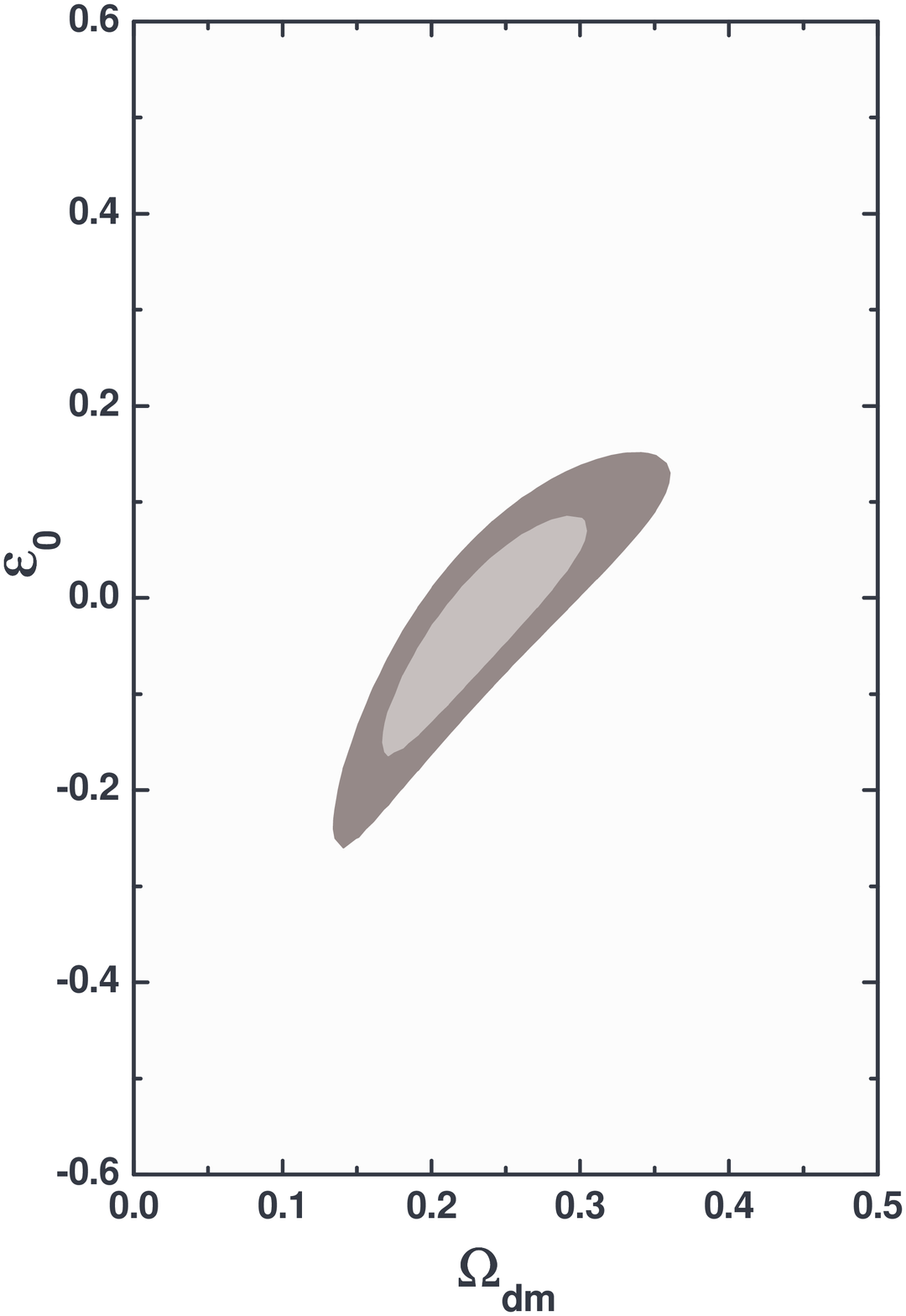,width=2.4truein,height=2.4truein,angle=0}}
\caption{The results of our statistical analyses. {\bf{a)}} Contours of $\chi^2$ in the plane $\epsilon_0 - \xi$. These contours are drawn for $\Delta \chi^2 = 2.30$ and $6.17$. Note that almost the entire interval of negative values of $\xi$ are ruled out at 2$\sigma$ unless $\epsilon_0 \simeq 0$ ($\Lambda$CDM model). {\bf{b)}} $\Delta \chi^2$ as a function of the interacting parameter $\epsilon_0$. From this analysis, we find $\epsilon = -0.08^{+0.18+0.33}_{-0.28-0.57}$ at 1 and 2$\sigma$ levels. {\bf{c)}} $\Omega_{dm,0} - \epsilon_0$ space for a constant interacting parameter $\epsilon = \epsilon_0$. At 1$\sigma$ level $\epsilon_0$ is restricted to the interval $\epsilon_0 = -0.03 \pm 0.03$.}
\label{fig:qzw}
\end{figure*}

\section{Observational analysis}

From now on we will discuss more quantitatively the observational viability of the class of interacting scenarios discussed above. To this end we perform a joint analysis involving current SNe Ia, BAO, CMB and $H_0$ data. Since we are particularly interested in bounds on the parameters $\epsilon_0$ and $\xi$ we fix $\Omega_{b,0} = 0.0416$ from WMAP results~\cite{cmbnew} (which is also in good agreement with the bounds on the baryonic component derived from primordial nucleosynthesis~\cite{nucleo}) and consider the recent determination of the Hubble parameter $H_0 = 74.2 \pm 4.8$~\cite{hubble} in conjunction with the CMB constraint $\Omega_{dm,0}h^2 = 0.109 \pm 0.006$~\cite{cmbnew}.
 
We use one of the most recent SNe Ia compilation, the so-called Union sample compiled in Ref.~\cite{union} which includes recent large samples from SNLS~\cite{snls} and ESSENCE~\cite{essence} surveys, older data sets and the recently extended data set of distant supernovae observed with the Hubble Space Telescope. The total compilation amounts to 414 SNe Ia events, which was reduced to 307 data points after selection cuts. 

We also use the distance ratio from $z_{\rm{BAO}} = 0.35$ to $z_{\rm{LS}} = 1089$, as measured by the Sloan Digital Sky Survey (SDSS), $\rm{R}_{\rm{BAO/LS}} = 0.0979 \pm 0.0036$~\cite{bao}. Here this quantity is given by
\begin{equation} \label{ratio}
{\rm{R}}_{{\rm{BAO/LS}}} =\frac{\left[\frac{z_{{\rm{BAO}}}}{E(z_{{\rm{BAO}}})}\right]^{1/3}r^{2/3}(z_{\rm{BAO}})}{r(z_{\rm{LS}})}\;,
\end{equation}
where $r(z) = \int_{0}^{z}{dz'}/{E(z')}$ is the comoving distance. Note that, although dark energy does not dominate early (see Figs. 1 and 2),  we do not use in our analysis either the BAO (${\cal{A}}$) or the CMB shift (${\cal{R}}$) parameters since these quantities use the approximation that the sound travel distance is $\propto 1/\sqrt{\Omega_m}$, which is not always true for interacting dark matter/energy  models due to the process of energy transfer (dark matter creation/annihilation~\cite{mca} or varying mass particles~\cite{vamp}) between these components (see, e.g., \cite{saulo} for a discussion). Note also that the above quantity seems to be slightly more precise than the ${\cal{A}}$ parameter since the scatter induced by uncertainties on $\Omega_{m}h^2$ cancels out in the ratio. In our analysis, therefore, we minimize the function $\chi^2 = \chi^{2}_{\rm{SNe}} + \chi^{2}_{\rm{R_{BAO/LS}}}$, which takes into account both the SNe Ia and BAO/CMB data discussed above (we refer the reader to Refs.~\cite{refs} for more on analyses involving different data sets).


The results of our statistical analyses are displayed in Fig. 5. Figure 5a shows confidence contours at 68.3\% and 95.4\% in the parametric space $\epsilon_0 - \xi$ that arise from the joint analysis described above. As expected, we note that the current observational bounds on $\xi$ are quite weak since it appears as a power of the scale factor in the energy density [Eqs. (\ref{dm}) and (\ref{decayv2})]. Note also that very large values of the interaction parameter $\epsilon_0$ are completely excluded regardless the value of $\xi$, and that the same is also true for almost the entire interval of negative values of $\xi$ unless $\epsilon_0 \simeq 0$ which, irrespective of the dimensionless parameter $\xi$, behaves very similarly to the standard $\Lambda$CDM model. 

To better visualize the constraints on $\epsilon_0$, in Fig. 5b we plot the plane $\epsilon_0 - \Delta \chi^2$. From this analysis we find $\epsilon_0 = -0.08^{+0.18}_{-0.28}$ at 1$\sigma$ level ($\chi^2_{min}/\nu = 1.02$), which means that both negative and positive values for the interacting parameter are allowed. Physically, this amounts to saying that not only is an energy flow from dark energy to dark matter ($\epsilon_0 > 0$) observationally allowed, but so is a flow from dark matter to dark energy ($\epsilon_0 < 0$) [see Eq. (\ref{dm})]. If we take in account the thermodynamical constraint derived in Ref.~\cite{alclim05}, i.e., $\epsilon_0 \geq 0$, we find $\epsilon_0 = 0.0^{+0.2}_{-0.0}$ at 1$\sigma$ level. 

For the sake of completeness, we also show in Fig. 5c the space $\Omega_{dm,0} - \epsilon_0$ for the case in which the interacting term is constant (i.e., $\xi = 0$)~\cite{wm,alclim05,ernandes}. Although physically more realistic and producing viable cosmic histories as shown in Figs. 1 and 2, clearly the introduction of a time-dependence on $\epsilon(a)$ (quantified by the parameter $\xi$) weakens the constraining power of the analysis. For the case $\epsilon=\rm{const.}$, the interacting parameter $\epsilon_0$ is more tightly bounded, i.e., $\epsilon_0 = -0.03 \pm 0.06$ at 95.4\% (C.L.). If we consider the constraint $\epsilon_0 \geq 0$~\cite{alclim05}, we find  $\epsilon_0 = 0.00^{+0.03 + 0.06}_{-0.00 -0.00}$ (68.3\% and 95.4\% C.L.).

\section{Final remarks}

We have discussed some cosmological consequences of an alternative mechanism of cosmic acceleration based on a general class of $\Lambda$-CDM interacting scenarios whose interaction term $\epsilon$ is deduced from the effect of the dark energy on the CDM expansion rate. We have gone a step further in the above description and extended the arguments of Refs.~\cite{wm,alclim05,ernandes} to a more realistic case in which $\epsilon$ is a function of time [see Eq. (\ref{Parametrization_a})]. The resulting expressions for the model are parameterized by the dimensionless parameters $\epsilon_0$ and $\xi$ and have many of the previous phenomenological approaches as a particular case. 

We have also investigated the dynamical behavior of these scenarios and found a number of viable cosmological solutions for a subset of the parameters $\epsilon_0$ and $\xi$ (Figs. 1 and 2). In particular, for large positive values of $\xi$ ($\gtrsim 0.8$) and $\epsilon_0 > 0$ we have found solutions of transient acceleration, in which the $\Lambda$-dark matter interaction will drive the Universe to a new matter-dominated era in the future. As mentioned earlier, this kind of solution seems to be in agreement with theoretical constraints from String/M theories on the quintessence potential $V(\phi)$ or, equivalently, on the dark energy equation-of-state $w$, as discussed in Ref.~\cite{fischler}.

From the observational point of view, we have investigated the current bounds on the parameterization (\ref{Parametrization_a}) from recent data of SNe Ia (Union sample), the distance ratio from baryon acoustic oscillation at $z_{\rm{BAO}} = 0.35$ and CMB decoupling at $z_{\rm{LS}} = 1089$, and $H_0$ estimates. We have shown that negative and positive values for the interacting parameter $\epsilon_0$ are observationally allowed, which means that both an energy flow from dark energy to dark matter as well as a flow from dark matter to dark energy are possible. For the parameter $\xi$ we have also found that positive values are largely favoured over negative ones. This includes all the well-behaved cases shown in Figs. 1 and 2.

Finally, it is worth emphasizing that we have restricted the present analysis to coupled quintessence models in which $w_\phi = -1$ (dynamical $\Lambda$ models) whereas a full treatment of the dark matter-dark energy interaction must also take into account the role of the dark energy equation-of-state in the process. Some theoretical and observational consequences of a $w$-CDM interacting scenario with a time-dependent coupling term, as well as a scalar field description for this class of models will appear in a forthcoming communication~\cite{new}.

\begin{acknowledgements}
FEMC acknowledges financial support from CAPES. JSA thank CNPq for the grants under which this work was carried out.

\end{acknowledgements}

\end{document}